\documentclass[twocolumn,showpacs,aps,epsfig]{revtex4}

\usepackage{graphicx}
\usepackage{epstopdf}
\usepackage{latexsym}

\usepackage[center]{subfigure}

\begin{document}

 \newcommand{\bq}{\begin{equation}}
 \newcommand{\eq}{\end{equation}}
 \newcommand{\bqn}{\begin{eqnarray}}
 \newcommand{\eqn}{\end{eqnarray}}
 \newcommand{\nb}{\nonumber}
 \newcommand{\lb}{\label}
\newcommand{\PRL}{Phys. Rev. Lett.}
\newcommand{\PL}{Phys. Lett.}
\newcommand{\PR}{Phys. Rev.}
\newcommand{\CQG}{Class. Quantum Grav.}

\title{Singularities  in Horava-Lifshitz theory }

\author{Rong-Gen Cai $^{a}$}
\email{cairg@itp.ac.cn}

\author{Anzhong Wang $^{b}$}
\email{anzhong_wang@baylor.edu}

\affiliation{$^{a}$ Key Laboratory of Frontiers in Theoretical  Physics, 
Institute of Theoretical Physics, Chinese Academy of Sciences, 
P.O. Box 2735, Beijing 100190, China\\
$^{b}$  GCAP-CASPER, Physics Department, Baylor
University, Waco, TX 76798-7316, USA }

\date{\today}

\begin{abstract}

Singularities in $(3+1)$-dimensional  Horava-Lifshitz (HL) theory of gravity  are studied.  These 
singularities can be divided into scalar, non-scalar curvature, and   coordinate singularities. Because of 
the  foliation-preserving diffeomorphisms of the theory, the  number of scalars that can be constructed 
from the extrinsic curvature tensor $K_{ij}$, the 3-dimensional Riemann tensor and their derivatives is 
much large than that constructed from the 4-dimesnional Riemann tensor and its derivatives in general 
relativity (GR). As a result, even for the same spacetime, it may be singular in the HL theory but not in GR. 
Two representative families of solutions with projectability condition are studied, one is the (anti-) de Sitter 
Schwarzschild solutions, and the other is the Lu-Mei-Pope (LMP) solutions written in a form satisfying the 
projectability condition - the generalized LMP solutions. The  (anti-) de Sitter  Schwarzschild solutions are 
vacuum solutions of both HL theory and GR, while the  LMP solutions with projectability condition satisfy
the HL equations coupled with an  anisotropic fluid with heat flow. It is found that  the scalars $K$ and 
$K_{ij}K^{ij}$ are singular  only at the center for the de Sitter  Schwarzschild solution, but singular at 
both the center and $ r = (3M/|\Lambda|)^{1/3}$ for the anti-de Sitter  Schwarzschild solution. The 
singularity at $ r = (3M/|\Lambda|)^{1/3}$ is absent in GR. In addition,  all the generalized LMP solutions 
have two scalar curvature singularities, located    at  either $r = 0$ and $r=r_{s} > 0$,  or $r=r_{1}$ and 
$r= r_{2}$ with $r_{2} > r_{1} > 0$,    or  $r=r_{s} > 0$ and $r = \infty$, depending on the choice of the 
free parameter $\lambda$.

\end{abstract}

\pacs{04.60.-m; 98.80.Cq; 98.80.-k; 98.80.Bp}

\maketitle

\section{Introduction}
\renewcommand{\theequation}{1.\arabic{equation}} \setcounter{equation}{0}

There has been considerable interest recently on a  theory of quantum gravity proposed 
by Horava \cite{Horava}, motivated by Lifshitz theory in solid state physics \cite{Lifshitz}, for which the 
theory is usually referred to as the Horava-Lifshitz (HL) theory.  The HL theory is based on the perspective 
that Lorentz symmetry should appear as an emergent symmetry at long distances, but can be fundamentally 
absent at high energies \cite{Pav}. With such a  perspective,  Horava considered systems whose scaling at short 
distances exhibits a strong anisotropy between space and time, 
\bq
\lb{1.1}
{\bf x} \rightarrow \ell {\bf x}, \;\;\;  t \rightarrow \ell^{z} t.
\eq
  In $(3+1)$-dimensional spacetimes, in order for the theory to be power-counting 
renormalizable,    it needs $z \ge 3$. At low energies, the theory is expected to  flow 
to $z = 1$, whereby the Lorentz invariance is ``accidentally restored." So, the HL theory is   non-relativistic, 
ultra-violet (UV) complete, explicitly breaks Lorentz invariance at short distances, but is expected to reduce 
to general relativity (GR) in the  infrared (IR) limit. 

The effective speed of light in this  theory diverges in the UV regime, which could potentially resolve the 
horizon problem without invoking inflation \cite{KK}. The  spatial curvature is enhanced by higher-order 
curvature terms \cite{calcagni,LMP,WM}, and this opens a new approach to the flatness problem and to 
a bouncing universe \cite{calcagni,brand,WW}. In addition,  in the super-horizon region scale-invariant  
curvature perturbations can be produced without inflation \cite{Muk,WMW}, and  the perturbations 
become adiabatic during slow-roll inflation driven by a single scalar field and the comoving curvature perturbation 
is constant  \cite{WMW}. Due to all  these remarkable features, the theory has attracted lot of  attention 
lately \cite{BHs,Cosmos,others}. 

To formulate the theory, Horava assumed two conditions -- {\em detailed balance and projectability} 
(He also considered the case where the detailed balance condition was  softly broken) \cite{Horava}. 
The detailed balance condition restricts the form of a general potential in a ($D+1$)-dimensional Lorentz action
to a specific form that can be expressed in terms of a D-dimensional action of a relativistic theory with Euclidean 
signature, whereby the number of independent-couplings is considerably limited. The projectability condition, on
the other hand, originates from the fundamental symmetry of the theory -- the foliation-preserving 
diffeomorphisms of the Arnowitt-Deser-Misner  (ADM) form,
 \bqn
 \lb{1.2}
ds^{2} &=& - N^{2}c^{2}dt^{2} + g_{ij}\left(dx^{i} + N^{i}dt\right)
     \left(dx^{j} + N^{j}dt\right), \nb\\
     & & ~~~~~~~~~~~~~~~~~~~~~~~~~~~~~~  (i, \; j = 1, 2, 3),~~~
 \eqn
which  require   coordinate transformations be only of the types,  
\bq
\lb{1.3}
t \rightarrow f(t),\; \;\; x^{i} \rightarrow \zeta^{i}(t, {\bf x}),
\eq
that is, space-dependent time reparameterizations are no longer 
allowed, although spatial diffeomorphisms are still a symmetry. 
Then, it is natural, but not necessary, to restrict the lapse function $N$
 to be space-independent, while the shift vector $N^{i}$ and 
the 3-dimensional metric $g_{ij}$ in general depend on both time and space,
\bq
\lb{1.4}
N = N(t), \;\;\; N^{i} = N^{i}(t, x),\;\;\; g_{ij} = g_{ij}(t, x).
\eq
This is the projectability condition, and  clearly is preserved by the foliation-preserving  diffeomorphisms
(\ref{1.3}). However, due to these restricted diffeomorphisms, one more degree of freedom appears
in the gravitational sector - a spin-0 graviton. This is potentially dangerous, and needs to be highly
suppressed in the IR regime, in order to be consistent with observations. Similar problems also raise 
in other modified theories, such as massive gravity \cite{RT}. 

Under the rescaling (\ref{1.1}), the dynamical variables $N, \; N^{i}$ and $g_{ij}$ scale as, 
 \bq
 \lb{1.5}
  N \rightarrow  N ,\;  N^{i}
\rightarrow {\ell}^{-2} N^{i},\; g_{ij} \rightarrow g_{ij}.
 \eq
Note that in \cite{WM,WMW}, the constant $c$ in the metric (\ref{1.2})  was absorbed into $N$, so that
there the lapse function scaled as $ \ell^{-2}$. 

So far most of the work on the HL theory has abandoned the projectability condition but  kept the detailed 
balance \cite{KK,LMP,BHs,Cosmos,others}. One of the main reasons is that the detailed balance condition 
leads to a very simple action, and the resulted theory is much  easier to deal with, while abandoning 
projectability condition gives rise to local rather than global Hamiltonian  constraint and energy conservation. 
However, with detailed  
balance a scalar field is not UV stable \cite{calcagni}, and gravitational perturbations in the scalar section
have ghosts  \cite{Horava} and are not stable for any given value of the dynamical coupling constant
$\lambda$ \cite{BS}.   In addition, detailed balance also requires a non-zero (negative) 
cosmological constant, breaks the parity in the purely gravitational sector \cite{SVW}, and makes the 
perturbations not scale-invariant \cite{GWBR}. Breaking the  projectability condition, on the other hand, 
can cause strong couplings  \cite{CNPS} and gives rise to an inconsistency theory \cite{LP}. 

To resolve these problems, various modifications have been proposed \cite{NewHLs}. In particular, 
Blas, Pujolas and Sibiryakov (BPS) \cite{BPS} showed that the strong coupling problem can be solved  
without projectability condition (in which the lapse function becomes dependent on both $t$ and $x^{i}$), 
when terms constructed from the  3-vector
\bq
\lb{1.5a}
a_{i} \equiv \frac{\partial_{i}N}{N},
\eq
are included. Contrary claims can be found in \cite{PS}. In addition, it is not clear how the inconsistency
problem \cite{LP} is resolved in such a generalization. 

On the other hand, Sotiriou, Visser and Weifurtner (SVW) formulated the most general HL theory with 
projectability but without detailed balance conditions \cite{SVW}.
The total action consists of three parts, kinetic, potential and matter,
 \bqn \lb{1.6}
S = \zeta^2\int dt d^{3}x N \sqrt{g} \left({\cal{L}}_{K} -
{\cal{L}}_{{V}}+\zeta^{-2} {\cal{L}}_{M} \right),
 \eqn
where $g={\rm det}\,g_{ij}$, 
and
 \bqn \lb{1.7}
{\cal{L}}_{K} &=& \frac{1}{c^{2}}\Big[K_{ij}K^{ij} - \left(1-\xi\right)  K^{2}\Big],\nb\\
{\cal{L}}_{{V}} &=& 2\Lambda - R + \frac{1}{\zeta^{2}}
\left(g_{2}R^{2} +  g_{3}  R_{ij}R^{ij}\right)\nb\\
& & + \frac{1}{\zeta^{4}} \left(g_{4}R^{3} +  g_{5}  R\;
R_{ij}R^{ij}
+   g_{6}  R^{i}_{j} R^{j}_{k} R^{k}_{i} \right)\nb\\
& & + \frac{1}{\zeta^{4}} \left[g_{7}R\nabla^{2}R +  g_{8}
\left(\nabla_{i}R_{jk}\right)
\left(\nabla^{i}R^{jk}\right)\right].
 \eqn
Here $\zeta^{2} = 1/{16\pi G}$, and $c$ denotes the speed of light. In the ``physical" units, one can set
$c = 1$ \cite{SVW}.  
The covariant derivatives and
Ricci and Riemann terms are all constructed from the three-metric $g_{ij}$,
while $K_{ij}$ is the extrinsic curvature,
 \bq \lb{1.8}
K_{ij} = \frac{1}{2N}\left(- \dot{g}_{ij} + \nabla_{i}N_{j} +
\nabla_{j}N_{i}\right),
 \eq
where $N_{i} = g_{ij}N^{j}$. The constants $\xi, g_{I}\,
(I=2,\dots 8)$  are coupling constants, and $\Lambda$ is the
cosmological constant. In the IR limit, all the high order curvature terms (with
coefficients $g_I$) drop out, and the total action reduces when
$\xi = 0$ to the Einstein-Hilbert action.

The SVW generalization  seems to have  the potential to solve  
the above mentioned problems \cite{Muka}, although it was found that gravitational scalar perturbations 
either have  ghosts ($0 \le \xi \le 2/3$) or are not  stable ($\xi < 0$) \cite{WM,Koz}. 
In order to avoid ghost instability, one needs to assume $\xi \le 0$. Then, the sound speed
$c_{\psi}^{2} = \xi/(2-3\xi)$ becomes imaginary, which leads to an IR instability. Izumi and Mukohyama 
showed that this type of instability does not show up if $|c_{\psi}|$ is less than a critical value \cite{IM}. 

It is fair to say, in order to have a viable HL theory, much work needs to be done, and various aspects of the 
theory ought to be explored, including the renormalization group flows \cite{IRS},  Vainshtein mechanism \cite{RT,Vain}, 
solar system tests \cite{IR},  Lorentz violations \cite{AGM}, and its applications to cosmology \cite{WM,WMW}.

In this paper, we shall study another important issue  in the HL theory - the problem of singularities, which is closely related
to the issue of black holes in this theory \cite{BHs}. Although we are initially interested in the case with projectability condition,
our conclusions can be equally applied to the HL theory without projectability condition. The extrinsic curvature $K_{ij}$ 
and the 3-dimensional Riemann tensor $R^{i}_{jkl}$ are not tensors under the  4-dimensional Lorentz transformations,
\bq
\lb{1.9}
x^{\mu} \rightarrow {x'}^{\mu} = \zeta^{\mu}\big(t, x^{i}\big).
\eq
As a result, in GR one usually does not use them to construct gauge-invariant quantities. However, in the HL theory,  due to the restricted 
diffeomorphisms, these quantities become tensors, and can be easily used to construct various  scalars.  If any of such scalars 
is singular, such a singularity cannot be limited by the restricted coordinate transformations  (\ref{1.3}). Then, we may say that the spacetime
is singular.  It is exactly in this vein that we 
study singularities in the HL theory.  In particular, we first generalize the definitions of scalar, non-scalar and coordinate singularities
in GR to the HL theory in Sec. II, and then in Sec. III we study two representative families of  spherical static solutions of the HL 
theory, and identify scalar curvature singularities using the three quantities $K, \; K_{ij}K^{ij}$ and $R$. 
In Sec. IV, we present our main conclusions and remarks. There is also an Appendix, in which we show explicitly that
the second class of the LMP solutions written in the ADM frame with projectability condition in general satisfy the HL equations coupled
with an anisotropic fluid with heat flow. 

Before proceeding further, we would like to note that black holes in GR for asymptotically-flat spacetimes are well-defined \cite{HE72}. 
However, how to generalize such definitions to more general spacetimes is still an open question \cite{Hay94,Wang}. The problem in 
the HL theory becomes more complicated \cite{KK2,IM}, partially because of the fact that particles in the HL theory can have non-standard 
dispersion relations, and therefore no uniform maximal speed exists. As a result, the notion of a horizon is observer-depedent.

\section{Singularities in HL theory}

\renewcommand{\theequation}{2.\arabic{equation}} \setcounter{equation}{0}

In GR, there are powerful Hawking-Penrose theorems \cite{HE72}, from which one can see that spacetimes
with quite ``physically reasonable" conditions are singular. Although the theorems did not tell the nature of the 
singularities, Penrose's cosmic censorship conjecture states that   those formed from 
gravitational collapse in a ``physically reasonable" situation are always covered by horizons \cite{Penrose}. 

To study further the nature of singularities in GR, Ellis and Schmidt  divided them into two different kinds, 
{\em spacetime curvature singularities} and {\em coordinate
singularities} \cite{ES77}. The former is real and cannot be made disappear by any  Lorentz transformations
(\ref{1.9}), while the latter is coordinate-depedent, and can be made disappear by proper Lorentz transformations.
 Spacetime curvature singularities are further divided into two sub-classes, {\em scalar curvature singularities} and 
{\em non-scalar curvature singularities}. If any of the scalars constructed from the 4-dimensional Riemann tensor 
$R^{\sigma}_{\mu\nu\lambda}$ and its derivatives
is singular, then the spacetime is said singular, and the corresponding singularity is a scalar one. If none of these scalars
is singular, spacetimes can be still singular. In particular,  tidal forces and/or  distortions 
(which are the double integrals  of the tidal forces), experienced by an observer, may become infinitely large 
 \cite{HWW02}. These kinds of singularities are usually referred to as non-scalar curvature singularities.

To generalize these definitions to the HL theory, as mentioned in the Introduction, both the extrinsic curvature $K_{ij}$
and the 3-dimensional Riemann tensor $R^{i}_{jkl}$ can be used to construct gauge-invariant quantities, as now they
are all tensors under the restricted transformations (\ref{1.3}). Then, we can see that now  there are three kinds of scalars: 
one is constructed
from  $K_{ij}$ and its derivatives;  one is from the 3-dimensional Riemann tensor $R^{i}_{jkl}$ and 
its derivatives; and  the other is the mixture of $K_{ij}, \; R^{i}_{jkl}$ and their derivatives. Therefore, we 
may define  a scalar curvature singularity in the HL theory as the one where any of   
the scalars constructed from $K_{ij}, \; R^{i}_{jkl}$ and their derivatives is singular.  A non-scalar curvature singularity
is the one where none of these scalars is singular, but some other physical quantities, such as tidal forces and distortions
experienced by observers, become unbounded. Coordinate singularities are the ones that can be limited by the
restricted coordinate transformations (\ref{1.3}). 

Several comments now are in order. The scalars constructed from the 3-dimensional tensors
$K_{ij}$ and $R^{i}_{jkl}$ and their derivatives include all scalars constructed from the 4-dimensional 
$R^{\sigma}_{\mu\nu\lambda}$ and its derivatives. Thus, according to the above 
definitions, all scalar singularities under the general Lorentz transformations (\ref{1.9}) are also scalar singularities
under  the restricted  transformations (\ref{1.3}), but not the other way around. In this sense, scalar singularities in
the HL theory are more general than those in GR. One simple example is the anti-de Sitter Schwarzschild solutions, 
which are also solutions of the SVW generalization with $\xi = 0$, as in this case the 3-dimensional Ricci tensor 
$R_{ij}$ vanishes identically, and the contributions of high order derivatives of curvature to the potential ${\cal{L}}_{V}$ 
are zero, as can be seen from Eq. (\ref{1.7}). However, as shown in the next section,   the corresponding two 
scalars $K$ and $K_{ij}K^{ij}$ all become singular at $r = (3M/|\Lambda|)^{1/3}$. This singularity is absent in
GR \cite{HE72}.

In 3-dimensional space,   the Weyl tensor vanishes identically, and the Riemann tensor is 
determined algebraically by the curvature scalar and the Ricci tensor: 
\bqn
\lb{2.1}
R_{ijkl} &=& g_{ik}R_{jl} + g_{jl}R_{ik} - g_{jk}R_{il} - g_{il} R_{jk}\nb\\
& &  - \frac{1}{2}\big(g_{ik}g_{jl} - g_{il}g_{jk}\big)R. 
\eqn
Therefore, the singular behavior of the scalars made of the 3-dimensional Riemann tensor $R^{i}_{jkl}$ 
may well be represented by the 3-dimensional curvature scalar $R$.  

\section{ Singularities in Spherical Static Spacetimes}

\renewcommand{\theequation}{3.\arabic{equation}} \setcounter{equation}{0}

The metric of general spherically symmetric static spacetimes that preserve the ADM form of Eq. (\ref{1.2}) 
with the projectability condition can be cast in the form \cite{GPW},  
\bqn
\lb{3.1}
ds^{2} &=& - dt^{2} + e^{2\nu} \left(dr + e^{\mu - \nu} dt\right)^{2} 
 + r^{2}d\Omega^{2}, 
\eqn 
 where    
$ \mu = \mu(r), \; \nu = \nu(r)$.
Then,  we have
\bqn
\lb{3.2}
R &=&    \frac{2e^{-2\nu}}{r^{2}}\Big[2r \nu' - \big(1-e^{2\nu}\big)\Big],\nb\\
K &=&e^{\mu -\nu}\left(\mu' + \frac{2}{r^{2}}\right),\nb\\
K_{ij}K^{ij} &=&e^{2(\mu -\nu)}\left(\mu'^{2} + \frac{2}{r^{2}}\right),
\eqn
where $\nu' \equiv d\nu/dr$, etc. It is interesting to note that for the metric (\ref{3.1}) we have
\bq
\lb{3.3}
C_{ij} = 0 = \epsilon^{ijk}R_{il}\nabla_{j}R^{l}_{k},
\eq
where 
$C_{ij}$ is the Cotton tensor, defined as
\bq
\lb{3.4}
C_{ij} =  \epsilon^{ikl}\nabla_{k}\left(R^{j}_{l} - \frac{1}{4}R \delta^{j}_{l}\right).  
\eq
As a result, the HL theory with detailed balance \cite{Horava}
\bqn
\lb{HLd}
S_{HLd} &=& \int{dt dx^{3} N \sqrt{g}\Bigg\{\frac{2}{\kappa^{2}}\Big[K_{ij}K^{ij} - \lambda  K^{2}\Big] }\nb\\
& & 
+ \frac{\kappa^{2}\mu^{2}\big(\Lambda_{W}R - 3\Lambda^{2}_{W}\big)}{8(1-3\lambda)}
+ \frac{\kappa^{2}\mu^{2}\big(1- 4\lambda) R^{2}}{32(1-3\lambda)} \nb\\
& & - \frac{\kappa^{2}\mu^{2}}{8}R_{ij}R^{ij} + \frac{\kappa^{2}\mu}{2w^{2}}\epsilon^{ijk}R_{il}\nabla_{j}R^{l}_{k}\nb\\
& & 
   - \frac{\kappa^{2}}{2w^{2}}C_{ij}C^{ij}\Bigg\}, 
\eqn
can be effectively considered as a particular case of the general SVW action (\ref{1.6}) with
\bqn
\lb{HLdr}
G &=& \frac{\kappa^{2}}{32\pi c^{2}},\;\;\;
c^{2} = \frac{\kappa^{4}\mu^{2}\Lambda_{W}}{16(1 - 3\lambda)},\;\;\; \Lambda = \frac{3\Lambda_{W}}{2}, \nb\\
g_{2} &=& \frac{4\lambda - 1}{4\Lambda_{W}}{\zeta^{2}},\;\;\;
g_{3} = \frac{1-3\lambda}{\Lambda_{W}}{\zeta^{2}},\nb\\
\xi &=& 1 - \lambda, \;\;\; g_{4} = g_{5} = ... = g_{8}  = 0.
\eqn
It should be noted that these relations are valid only when the conditions of Eq. (\ref{3.3}) hold. In general,
these two terms do not vanish and violate parity, while the SVW action always preserves it. It must not be confused with the parameter $\mu$ used
in the action (\ref{HLd}) and the metric coefficient used in (\ref{3.1}).

To study singularities in the HL theory, in the rest of this paper we shall restrict ourselves to two representative cases, the 
(anti-) de Sitter Schwarzschild solutions and the solutions found by Lu, Mei and Pope (LMP) \cite{LMP}. 

\subsection{(Anti-) de Sitter Schwarzschild Solutions }

The (anti-) de Sitter Schwarzschild solutions are given by \cite{GPW}
\bq
\lb{3.5}
\mu = \frac{1}{2}\ln\left(\frac{M}{r} + \frac{\Lambda}{3}r^{2}\right),\;\;\;
\nu = 0.
\eq
When $\Lambda > 0$, it represents  the de Sitter Schwarzschild solutions, and when $\Lambda < 0$ it represents  the 
anti-de Sitter Schwarzschild solutions. As mentioned previously, they are also solutions of the SVW generalization with $\xi = 0$.
Inserting the above into Eq.(\ref{3.2}), we find that 
\bqn
\lb{3.6}
R &=&   0,\nb\\
K &=& \left(\frac{3M + \Lambda r^{3}}{12 r^{3}}\right)^{1/2} \left(\frac{4}{r} - \frac{3M -  2\Lambda r^{3}}{3M + \Lambda r^{3}}\right),\nb\\
K_{ij}K^{ij} &=& \frac{3M + \Lambda r^{3}}{12 r^{3}}\left[8 + \left(\frac{3M -  2\Lambda r^{3}}{3M + \Lambda r^{3}}\right)^{2}\right].
\eqn
Clearly, when $\Lambda \ge  0$, $K$ and $K_{ij}K^{ij}$ are singular only at the center $r = 0$. However, when $\Lambda < 0$, they
are also singular   at $r = r_{\Lambda} \equiv (3M/|\Lambda|)^{1/3}$. In contrast to GR, this singularity is 
a scalar one, and cannot be removed by any coordinate transformations given by Eq. (\ref{1.3}). 

In GR, the (anti-) de Sitter Schwarzschild solutions are usually given in the orthogonal gauge,
\bq
\lb{3.7}
ds^{2} =  - e^{2\Psi(r)}d\tau^{2}  + e^{2\Phi(r)}dr^{2} + r^{2}d\Omega^{2},
\eq
with  
\bq
\lb{3.8}
\Psi = -\Phi =  \frac{1}{2}\ln\left(1 - \frac{2M}{r} + \frac{1}{3}\Lambda r^{2}\right).
\eq
Clearly, the metric (\ref{3.7}) does not satisfy the projectability condition, and its coefficient $g_{rr}$ is singular at   $e^{2\Phi(r_{EH})} = 0$.
But, in GR this is a coordinate singularity, and all scalars made of the 4-dimensional Riemann tensor  and its derivatives are finite at $r = r_{EH}$.

It is interesting to note that in the orthogonal gauge (\ref{3.7}), $K$ and $K_{ij}K^{ij}$  all vanish, as can be seen from Eqs. (\ref{1.8}), while the 
3-dimensional Ricci scalar is given by $R_{orth} = - 2\Lambda$, where $R_{orth}$ denotes the quantity calculated in the 
orthogonal gauge (\ref{3.7}). 

In \cite{GPW}, it was showed explicitly that metric (\ref{3.7}) is related to metric (\ref{3.1}) by the coordinate transformations,
\bq
\lb{3.10}
\tau = t -     \int^{r}{ \sqrt{e^{-2\Psi} - 1}\; e^{\Phi} dr},
\eq
under which we have
\bqn
\lb{3.11a}
\Phi(r) &=&  \nu(r) - \frac{1}{2}\ln\Big(1 -  e^{2\mu}\Big), \nb\\  
\Psi(r) &=& \frac{1}{2}\ln\Big(1 - e^{2\mu}\Big),
\eqn
or inversely,
\bq
\lb{3.11b}
 \mu = \frac{1}{2}\ln\Big(1 -  e^{2\Psi}\Big), \;\;\;
\nu = \Phi(r) + \Psi(r).
\eq
Note that   the coordinate transformations (\ref{3.10}) are not allowed by the  foliation-preserving 
diffeomorphisms (\ref{1.3}). In addition, since $K,\; K_{ij}K^{ij}$ and $R$ are not scalars under  these transformations,  
it explains why we have completely different physical interpretations  in the two  different gauges, defined, respectively, 
by Eqs. (\ref{3.1}) and (\ref{3.10}). 
In fact, in the framework of the HL theory the two different gauges represent 
 two different theories - metric (\ref{3.1}) represents a HL theory with projectability condition, while metric
 (\ref{3.10}) represents a HL theory without projectability condition.
 
 \subsection{ The LMP Solutions }
 
\begin{figure}
\includegraphics[width=\columnwidth]{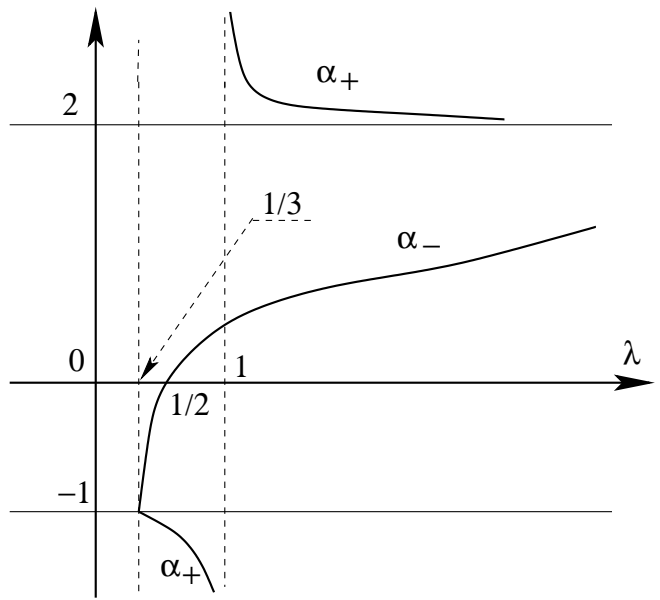}
\caption{The functions $\alpha_{\pm}(\lambda)$ defined by Eq. (\ref{3.13}).  }
\label{fig1}
\end{figure}

 The LMP solutions were originally found for the HL theory with detailed balance   but 
 without projectability conditions in the form (\ref{3.7}). There are two classes of solutions, given, respectively,
 by
 \bq
 \lb{3.12a}
 \Phi = - \frac{1}{2}\ln\big(1  + x^{2}\big),
 \eq
 for any $\Psi(r)$, and 
 \bqn
 \lb{3.12b}
  \Phi &=& - \frac{1}{2}\ln\Big(1 + x^{2} - \alpha x^{\alpha_{\pm}}\Big),\nb\\
  \Psi &=&- \beta_{\pm}\ln(x)  + \frac{1}{2}\ln\Big(1 + x^{2} - \alpha x^{\alpha_{\pm}}\Big),
 \eqn
 where $x \equiv \sqrt{-\Lambda_{W}}\; r$, and
 \bqn
 \lb{3.13}
  \alpha_{-} &\equiv& \frac{2\lambda - \sqrt{6\lambda -2}}{\lambda -1} =  \frac{2(2\lambda -1)}{2\lambda + \sqrt{6\lambda -2}},\nb\\
 \alpha_{+} &\equiv& \frac{2\lambda + \sqrt{6\lambda -2}}{\lambda -1}, \;\;\; 
\beta_{\pm} \equiv 2\alpha_{\pm} -1,
  \eqn
with $\alpha$ being an arbitrary real constant. Fig. \ref{fig1} schematically shows the curves of $\alpha_{\pm}$ vs $\lambda$, which shows
that there is no solution for the $\alpha_{+}$ branch when $\lambda = 1$.

 Solutions given by Eqs. (\ref{3.12a}) and (\ref{3.12b}) shall be referred to, respectively,  as Class A and B solutions.
Particular values of $\alpha_{\pm}$ are
\bqn
 \lb{3.13a}
  \alpha_{-}(\lambda)  &=& \cases{-1, & $\lambda = 1/3$,\cr
  0, & $\lambda = 1/2$,\cr
  1/2, & $\lambda = 1$,\cr
  2/3, & $\lambda \simeq 1.375$,\cr
  1, & $\lambda = 3$,\cr
  2, & $\lambda  = \infty$,\cr} \nb\\
 \alpha_{+}(\lambda)   
 &=& \cases{-1, & $\lambda = 1/3$,\cr
  -\infty, & $\lambda = 1^{-0}$,\cr
  +\infty, & $\lambda = 1^{+0}$,\cr 
    2, & $\lambda  = \infty$,\cr}  
  \eqn
which are useful in the following discussions.

In the orthogonal gauge (\ref{3.7}), as mentioned previously, the extrinsic curvature $K_{ij}$ vanishes identically, while the 3-dimensional
curvature for the above two classes of solutions can be written as 
\bq
\lb{3.14}
R_{orth} = \frac{2}{r^{2}}\Big\{\alpha\big(1 + \alpha_{\pm}\big)x^{\alpha_{\pm}} - 3x^{2}\Big\}.
\eq
When $\alpha = 0$, the corresponding $R_{orth}$ is for the solution (\ref{3.12a}), which reduces to a constant. That is, in this case all the three scalars
$K,\; K_{ij}K^{ij}$ and $R$ are finite. When $\alpha \not= 0$,  it is for the solution (\ref{3.12b}), which is singular only when $\alpha_{\pm} \le 2$ at $r = 0$. 
From Fig.\ref{fig1} we can see that $\alpha_{-}$ is always less than two for a finite $\lambda$. Therefore, this branch of solutions is always singular 
at the center. $\alpha_{+}$ is always less than two for $1/3 \le \lambda < 1$ and greater than two for $\lambda > 1$. Therefore, $R_{orth}$ is finite 
for  the $\alpha_{+}$ solutions with $\lambda > 1$ for any $r$, including the center $r = 0$, while it is singular at the center for the $\alpha_{+}$ solutions
 with  $1/3 \le \lambda < 1$.

 \subsubsection{Class A Solutions}
 
 Transforming the above solutions into the canonical ADM form (\ref{3.1}), we find that for the solution (\ref{3.12a}), we have
 \bq
 \lb{3.15}
 \mu = - \infty,\;\;\; \nu = - \frac{1}{2}\ln\Big(1 - \Lambda_{W} r^{2}\Big),
 \eq
 where in writing the above expressions, we had chosen $\Psi = 0$. Then, the corresponding three scalars are given by
 \bq
 \lb{3.16}
 K = K_{ij}K^{ij} = 0, \;\;\; R = 6 \Lambda_{W},
 \eq
 which are all finite. As shown in \cite{GPW}, this solution is also a vacuum solution of the SVW generalization  (\ref{1.6}) with a non-vanishing 
 constant curvature $k = 6 \Lambda_{W}$. Since it does not depend explicitly on the coupling constant $\xi$, it is a vacuum solution for
 any given $\xi$, including $\xi = 0$. For detail, we refer readers to Sec. V of \cite{GPW}.  
 
 \subsubsection{Class B Solutions}
 
 For Class B solutions (\ref{3.12b}), they can be written in the canonical ADM form (\ref{3.1}) with
 \bq
 \lb{3.16}
 \mu = \frac{1}{2}\ln\Delta,\;\;\;
 \nu = \big(1 - 2\alpha_{\pm}\big)\ln(x),
 \eq
 where
 \bq
 \lb{3.17}
 \Delta \equiv 1 - x^{2(1 - 2\alpha_{\pm})} - x^{4(1 - \alpha_{\pm})} + \alpha x^{2 - 3\alpha_{\pm}}.
 \eq
 Clearly, to have real solutions, we must assume $\Delta \ge 0$. 
  It should be noted that unlike Class A solutions, this class of solutions
 do not satisfy the vacuum equations of the HL theory with projectability condition, due to the fact that the HL actions 
 are not invariant under the coordinate transformations (\ref{3.10}). As shown in Appendix, they can be interpreted as 
 solutions of the HL theory with projectability condition coupled with a spherical anisotropic fluid with heat flow.  The
properties of singularities of these quantities can be well represented by the three scalars $K,\; K_{ij}K^{ij}$ and $R$, 
as can be seen from Appendix, so in the following we shall not consider them specifically.

 Inserting the above solutions into Eq. (\ref{3.2}) we find that
 \bqn
 \lb{3.18}
 R &=& \frac{2}{r^{2}}\Big\{1 - \big(1 + 2\beta_{\pm}\big)x^{2\beta_{\pm}}\Big\},\nb\\
 K &=&  \frac{\sqrt{-\Lambda_{W}}}{2\Delta^{1/2} x^{3 - 2\alpha_{\pm}}}\left(4 \sqrt{-\Lambda_{W}} \Delta - x^{2}\delta\right),\nb\\
 K_{ij}K^{ij} &=&  - \frac{\Lambda_{W}}{4\Delta x^{4(1 - \alpha_{\pm})}}\left(8\Delta - x^{2}\delta^{2}\right),
 \eqn
 where 
 \bqn
 \lb{3.19}
 \delta &\equiv& 2\big(1-2\alpha_{\pm}\big)x^{1-4\alpha_{\pm}} + 4\big(1-\alpha_{\pm}\big)x^{3-4\alpha_{\pm}}\nb\\
 & & ~~~
                     - \alpha\big(2-3\alpha_{\pm}\big)x^{1-3\alpha_{\pm}}.
 \eqn
To study these solutions further, it is found convenient to consider the two branches  $\alpha_{\pm}$ separately. 
We shall use $\beta = \beta_{\pm}$ to denote the $\alpha_{\pm}$ branches. 

\begin{figure}
\includegraphics[width=\columnwidth]{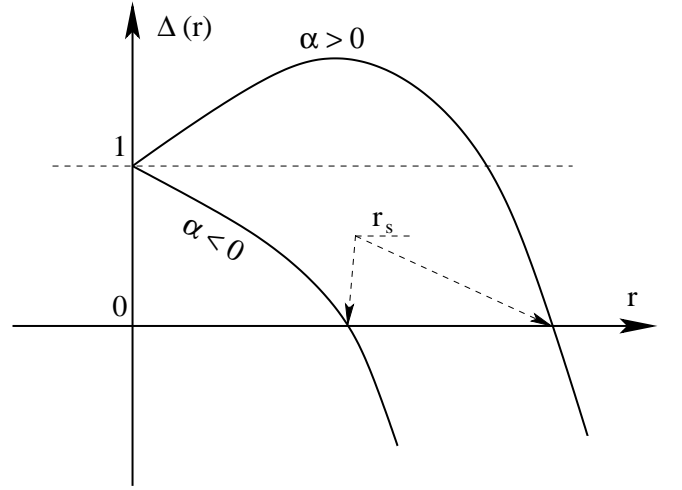}
\caption{The function $\Delta(x)$ defined by Eq. (\ref{3.17}) for the $\alpha_{+}$ branch with $1/3 \le \lambda < 1$. All
three scalars  $R, \; K$ and $K_{ij}K^{ij}$, defined by Eq. (\ref{3.18}), become unbounded at $r = 0$, while only $K$ 
and $K_{ij}K^{ij}$ diverge at $r = r_{s}$. }
\label{fig2}
\end{figure}

{\bf Case i) $\beta = \beta_{+},\; 1/3 \le \lambda < 1$:} In this case we have  $\alpha_{+}\le -1 $ and $\beta_{+} \le -3$. 
Then, from the expression (\ref{3.17}) we find that
\bq
\lb{3.20a}
\Delta = \cases{1, & $x = 0$,\cr
- \infty, & $ x \rightarrow \infty$.\cr}
\eq
Thus, there must exist a point $x= x_{s}$ at which we have $\Delta(x_{s}) = 0$. Since $\Delta \ge 0$, we must restrict the 
solutions to
the range $0 \le x \le x_{s}$ (or equivalently, $ 0 \le r \le r_{s}$, where $r_{s} \equiv x_{s}/\sqrt{-\Lambda_{W}}$.). 
Fig. \ref{fig2} shows this case. From the expressions (\ref{3.18})
we find that all these three scalars diverge at the center $r = 0$, while $K$ and $K_{ij}K^{ij}$ diverge at $r = r_{s}$. Therefore,
in the present case there are two scalar curvature singularities, located, respectively, at $r = 0$ and $r = r_{s}$.
 
\begin{figure}
\includegraphics[width=\columnwidth]{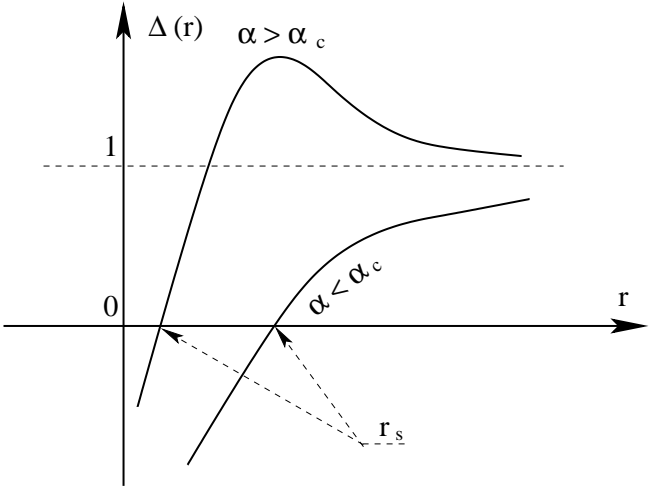}
\caption{The function $\Delta(x)$ defined by Eq. (\ref{3.17}) for the $\alpha_{+}$ branch with $  \lambda > 1$, where
$\alpha_{c}$ is the value of $\alpha$, for which $\Delta'(x) > 0$ for $\alpha < \alpha_{c}$. $K$ and $K_{ij}K^{ij}$ diverge 
at $r = r_{s}$, while $R$ diverges as $r \rightarrow \infty$.}
\label{fig3}
\end{figure}

{\bf Case ii) $\beta = \beta_{+},\;   \lambda > 1$:} In this case we have  $\alpha_{+} \ge 2 $ and $\beta_{+} \ge 3$, where the equality holds only
when $\lambda = \infty$. 
Then, from the expression (\ref{3.17}) we find that
\bq
\lb{3.20b}
\Delta = \cases{- \infty, & $x = 0$,\cr
+ 1, & $ x \rightarrow \infty$.\cr}
\eq
Fig.\ref{fig3} shows this case, from which we find that for any given $\lambda > 1$, there always exists a minimum $r_{s} > 0$ so that
$\Delta(r) \ge 0$ for $r \ge r_{s}$, where $r_{s}$ is the solution of $\Delta(r) = 0$.  Then, from the expressions (\ref{3.18})
we find that  $K$ and $K_{ij}K^{ij}$ diverge at $r = r_{s}$, while $R$ diverges as $r \rightarrow \infty$. Therefore,
in the present case there are also two scalar curvature singularities, located, respectively, at $r = r_{s}$ and $r = \infty$.
 
\begin{figure}
\includegraphics[width=\columnwidth]{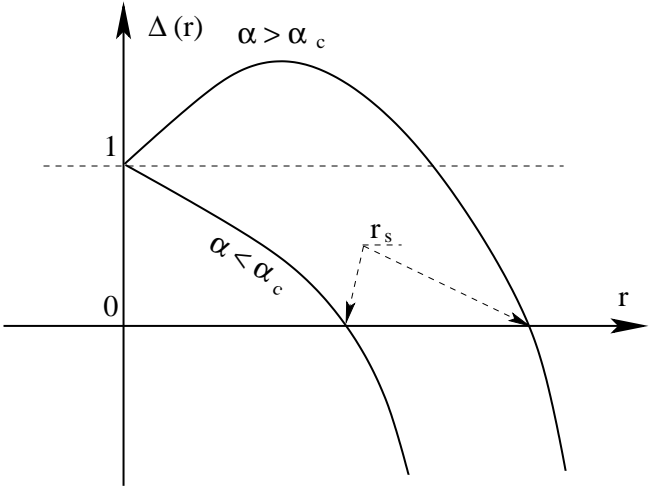}
\caption{The function $\Delta(x)$ defined by Eq. (\ref{3.17}) for the $\alpha_{-}$ branch with $1/3 \le  \lambda < 1$, where
$\alpha_{c}$ is the value of $\alpha$, for which $\Delta'(x) < 0$ for $\alpha < \alpha_{c}$. $R, \; K$ and $K_{ij}K^{ij}$ diverge 
at the center $r = 0$, while only $K$ and $K_{ij}K^{ij}$  diverges at $r = r_{s}$.}
\label{fig4}
\end{figure}

 {\bf Case iii) $\beta = \beta_{-},\;  1/3 \le  \lambda < 1$:} In this case we have  $-1 \le \alpha_{-} < 1/2 $ and $-3 \le \beta_{-} < 0$, 
 where the equality holds only for $\lambda = 1/3$. 
Then, from the expression (\ref{3.17}) we find that
\bq
\lb{3.20c}
\Delta = \cases{+1, & $x = 0$,\cr
- \infty, & $ x \rightarrow \infty$.\cr}
\eq
from Fig.\ref{fig4}, we find that for any given $\lambda$ in this range, there always exists a maximum $r_{s} > 0$ so that
$\Delta(r) \ge 0$ for $0\le r \le r_{s}$.  Then, From the expressions (\ref{3.18})
we find that  all three scalars $R, \; K$ and $K_{ij}K^{ij}$ become unbounded at the center, while only  $K$ and $K_{ij}K^{ij}$ diverge at $r = r_{s}$. Therefore,
in the present case there are two scalar curvature singularities, located, respectively, at $r = 0$ and $r = r_{s}$. 
 
 {\bf Case iv) $\beta = \beta_{-} = 0,\;     \lambda  = 1$:} In this case we have $\alpha_{-} = 1/2$, and 
 \bq
\lb{3.20d}
\Delta = \alpha x^{1/2} - x^{2}.  
\eq
Thus,   to have $\Delta$ non-negative, we must assume $\alpha > 0$. Then, $\Delta \ge 0$ for $0 \le r \le  r_{s} \equiv \alpha^{2/3}$, where
$\Delta(x) = 0$ at both $r = 0$ and $r = r_{s}$. At the center, all three scalars $R, \; K$ and $K_{ij}K^{ij}$ become unbounded, while 
only  $K$ and $K_{ij}K^{ij}$ diverge at $r = r_{s}$.

\begin{figure}
\includegraphics[width=\columnwidth]{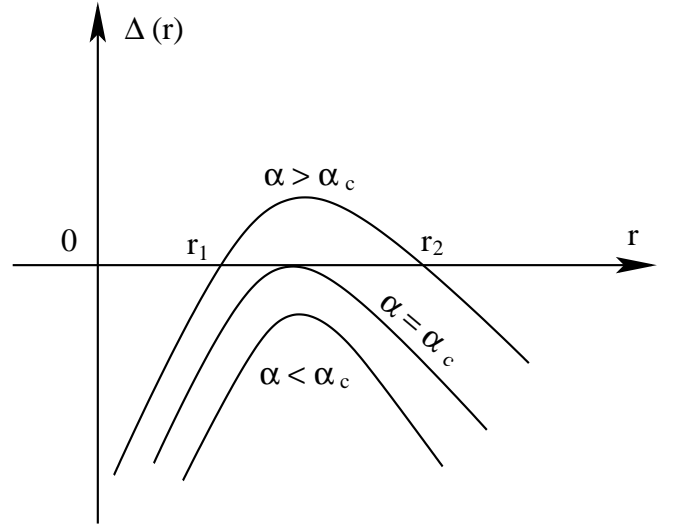}
\caption{The function $\Delta(x)$ defined by Eq. (\ref{3.17}) for the $\alpha_{-}$ branch with $1 <  \lambda < 3$, where
$\alpha_{c}$ is the value of $\alpha$, for which $\Delta(x) \ge 0$ is possible only when $\alpha > \alpha_{c}$. $K$ and $K_{ij}K^{ij}$ diverge 
at the two points $r = r_{1,2}$, while $R$  remains finite there.}
\label{fig5}
\end{figure}

 {\bf Case v) $\beta = \beta_{-},\;  1 <  \lambda < 3$:} In this case we have  $1/2 \le \alpha_{-} < 1,\; 0 \le \beta_{-} < 1$, and 
\bq
\lb{3.20e}
\Delta = \cases{-\infty, & $x = 0$,\cr
- \infty, & $ x \rightarrow \infty$.\cr}
\eq 
Fig.\ref{fig5} shows the general properties of $\Delta(r)$, from which we can see that for any given $\lambda$, there always exists a critical value
$\alpha_{c}$, for which $\Delta(r) \ge 0$ is possible only when $\alpha > \alpha_{c}$. In the latter case, $\Delta (r) = 0$ always has two positive roots,
say, $r_{1}$ and $r_{2}$. Without loss of generality, we assume  $r_{2} > r_{1} > 0$. When $r_{1} \le r \le r_{2}$, $\Delta (r) $ is non-negative.
At the two points $r= r_{1}$ and $r_{2}$ we have $\Delta(r_{i}) = 0$, and Eq. (\ref{3.18}) shows that both $K$ and $K_{ij}K^{ij}$ become unbounded 
at these points, while $R$ remains finite.    Therefore,
in the present case there are also two scalar curvature singularities, located, respectively, at $r = r_{1}$ and $r = r_{2}$ for $\alpha > \alpha_{c}$.
Solutions with $\alpha \le \alpha_{c}$ are not physical.
 
\begin{figure}
\includegraphics[width=\columnwidth]{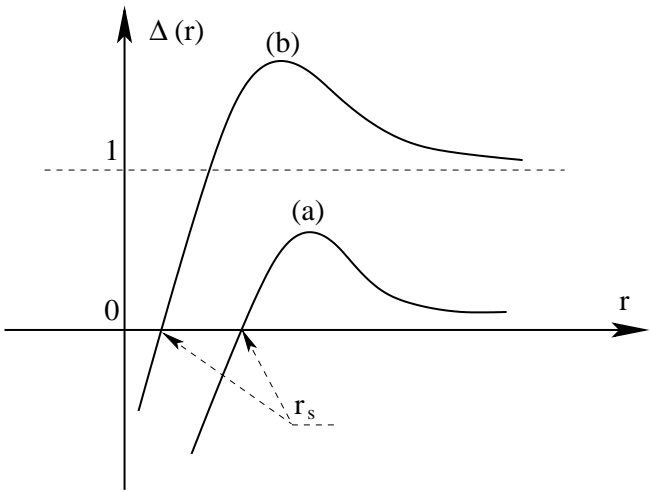}
\caption{The function $\Delta(x)$ defined by Eq. (\ref{3.17}) for the $\alpha_{-}$ branch with (a) $\lambda =  3,\; \alpha > 0$; and 
(b) $\lambda > 3$ for any $\alpha$. }
\label{fig6}
\end{figure}

  {\bf Case vi) $\beta = \beta_{-} = 1,\;    \lambda = 3$:} In this case we have  $  \alpha_{-} =  \beta_{-} = 1$, and 
\bq
\lb{3.20f}
\Delta =  \frac{\alpha}{x} - \frac{1}{x^{2}} = \cases{- \infty, & $ x = 0$,\cr
0, & $x \rightarrow \infty$.\cr}
\eq 
Clearly, to have $\Delta \ge 0$, we must assume  $\alpha > 0$. Then, for $r \ge r_{s} \equiv 1/(\sqrt{-\Lambda_{W}}\; \alpha)$,  $\Delta$ is non-negative
[See Fig.\ref{fig6}, Curve (a)]. 
At $r = r_{s}$, $\Delta(r) = 0$, and  Eq. (\ref{3.18}) shows that both $K$ and $K_{ij}K^{ij}$ become unbounded 
at this point, while $R$ remains finite.   As $r \rightarrow \infty$,  $\Delta(r) \rightarrow 0$, and $K$ and $K_{ij}K^{ij}$ become unbounded again,
while $R$ still remains finite. Therefore,
in the present case there are also two scalar curvature singularities, located, respectively, at $r =  r_{s}$ and $r = \infty$.

  {\bf Case vii) $\beta = \beta_{-},\;     \lambda > 3$:} In this case we have  $\alpha_{-} > 1,\; \beta_{-} > 1$, and 
\bq
\lb{3.20g}
\Delta = \cases{-\infty, & $x = 0$,\cr
+1, & $ x \rightarrow \infty$.\cr}
\eq 
Fig.\ref{fig6}(b) shows the general properties of $\Delta(r)$, from which we can see that for any given $\lambda$, there always exists a point
$r = r_{s}$ where  $\Delta(r_{s}) = 0$. When $r > r_{s}$,   $\Delta$ is always positive.   Eq. (\ref{3.18}) shows that now
both $K$ and $K_{ij}K^{ij}$ diverge at  $r = r_{s}$, while all three scalars become unbounded as $r \rightarrow \infty$.    Thus,
in the present case there are two scalar curvature singularities, located, respectively, at $r = r_{s}$ and $r = \infty$.
 
 \section{Conclusions}


\renewcommand{\theequation}{4.\arabic{equation}} \setcounter{equation}{0}

 In this paper, we have  studied singularities in the HL theory, and classified them into three different kinds, the scalar, non-scalar, 
 and coordinate  singularities, following the classification given in GR \cite{ES77}. Due to  the restricted 
 diffeomorphisms (\ref{1.3}),  the number of the scalars that can be constructed from the extrinsic curvature tensor $K_{ij}$, the 3-dimensional Riemann 
 tensor $R^{i}_{jkl}$ and their derivatives is much larger than that constructed from the 4-dimensional Riemann tensor 
 $R^{\sigma}_{\mu\nu\lambda}$ and its derivatives. The latter is invariant  under the general Lorentz transformations (\ref{1.9}). As a result, even for 
 the same spacetime, it may be singular in the HL theory, but not singular in GR. One simple example is the anti-de Sitter Schwarzschild 
 solution written in the ADM form (\ref{3.1}). This solution is a solution of both HL theory and GR. However,
 in the HL theory, there are two scalar singularities located, respectively, at the origin $r = 0$ and  $r = (3M/|\Lambda|)^{1/3}$, as
 the two scalars $K$ and $K_{ij}K^{ij}$ become singular at these points. It is well-known that the scalar singularity at $r = (3M/|\Lambda|)^{1/3}$
  is absent  in GR. This is because both $K$ and $K_{ij}K^{ij}$ are not scalars under the general Lorentz transformations (\ref{1.9}). Thus,
  even they are singular at this point, it only represents a coordinate singularity, and all  scalars constructed from  the 4-dimensional 
  Riemann tensor and its derivatives are finite. On the other hand, due to the restricted transformations (\ref{1.3}),  $K$ and $K_{ij}K^{ij}$ are
  scalars in the HL theory, and once they are singular, the resulting singularity cannot be transferred away by the restricted
  transformations. As a result, in the HL theory it represents  a real spacetime singularity. 
  
  With the above in mind, we have  studied the LMP solutions \cite{LMP}, and found that their singularity behavior in the orthogonal frame
  defined by (\ref{3.7}) is different from that in the ADM frame defined by (\ref{3.1}) for the second class of the LMP solutions. In particular,
   in the orthogonal frame, $K_{ij}$ vanishes,
  so do $K$ and  $K_{ij}K^{ij}$, while the 3-dimensional Ricci scalar $R$ [cf. Eq. (\ref{3.14})] can be singular at the origin or infinity, depending
  on the choice of the parameter $\lambda$. However, in the   ADM frame at least one of the three scalars $K, \; K_{ij}K^{ij}$ and $R$ is
  always singular at two different points, either   $r = 0$ and $r=r_{s} > 0$,  or $r=r_{1}$ and $r= r_{2}$
with $r_{2} > r_{1} > 0$, or  $r=r_{s} > 0$ and $r = \infty$, depending on the choice of the free 
parameter $\lambda$, where $r_{s}$ is a finite non-zero positive constant.   
This different singular behavior originates from the fact that the two frames are related
  by the coordinate transformations (\ref{3.10}), which is not allowed by the foliation-preserving diffeomorphisms (\ref{1.3}), or in other words,
  $K, \; K_{ij}K^{ij}$ and $R$ are not scalars under such  transformations. In fact, in the framework of the HL theory, the two frames actually 
  represent two different HL theories, one is with the projectability condition, while the other is without. In particular, the second class of the LMP
  solutions in the orthogonal frame satisfy the vacuum HL equations, while in the ADM frame they satisfy the HL equations coupled with an
  anisotropic fluid with heat flow, as shown explicitly in the Appendix.
  
  Our above results  show clearly that the problem of singularities in the HL theory  is a very delicate problem, due to the restricted 
 diffeomorphisms (\ref{1.3}), which preserve the ADM foliations (\ref{1.2}). Further investigations are needed, in particular, in terms of the strength 
 of these singularities.
 In the examples studied in this paper, all singularities indicated by the two scalars  $K$ and $K_{ij}K^{ij}$ at $r= r_{s} > 0$, including that of the
 anti-de Sitter Schwarzschild solution, seems weak in the sense of tidal forces and distortions experiencing by observers. Therefore, it is
 not clear whether or not  the spacetime  is extendable  across such a singularity \cite{HWW02}. 
 
 Finally, we would like to note that the ADM form (\ref{3.1}) can be considered as a particular case of the HL theory without projectability condition.
 Therefore, restricting ourselves only to the  HL theory without projectability condition does not solve the singularity problem occurring at $r = r_{s}$.

~\\{\bf Acknowledgements:} 

We would like to express our gratitude to  Jared Greenwald and Antonios  Papazoglou  for valuable discussions and suggestions.    
Part of the work  was supported   by NNSFC   under Grant 10535060, 10821504 and 10975168 (RGC); and  No. 10703005 and 
No. 10775119 (AZ).

 \section*{Appendix: The Generalized LMP Solutions with Projectability Condition}
 \renewcommand{\theequation}{A.\arabic{equation}} \setcounter{equation}{0}

 The second class of the LMP solutions with projectability condition takes the form of
 Eq. (\ref{3.1}) with $\mu$ and $\nu$ given by Eq. (\ref{3.16}). Unlike the first class, this
 one does not satisfy the HL vacuum equations with projectability condition, due to the
 fact that HL actions  are not invariant under the coordinate transformations (\ref{3.10}). In particular,
 under these  transformations, we have
 \bq
 \lb{a.1}
 R_{ij} \rightarrow R_{ij} + \delta{R}_{ij}.
 \eq
 For  spherically symmetric static solutions,  the extra term $\delta{R}_{ij}$ in general gives  rise to
 an anisotropic fluid with heat flow \cite{GPW}, possibly subjected to some energy conditions
 \cite{HE72}. In particular, 
 for the LMP solutions of Eqs. (\ref{3.1}) 
  and (\ref{3.16}), the energy density $J^{t}$, defined by
   \bq 
   \lb{a.2}
J^{t} = 2\left(N\frac{\delta{\cal{L}}_{M}}{\delta N} +
{\cal{L}}_{M}\right),
 \eq
is given by   the Hamiltonian equation,
 \bq 
  \lb{a.3}
\int{ d^{3}x\sqrt{g}\left({\cal{L}}_{K} + {\cal{L}}_{{V}}\right)}
= 8\pi G \int d^{3}x {\sqrt{g}\, J^{t}},
 \eq
where
\bqn
\lb{a.4}
{\cal{L}}_{K} &=& - \frac{\Lambda_{W} e^{2(\mu-\nu)}}{4 x^{2}\Delta}
\Big\{\xi x^{2}{\Delta'}^{2} - 8(1-\xi)x \Delta\Delta' \nb\\
& & - 8(1-2\xi)x \Delta^{2}\Big\},\nb\\
{\cal{L}}_{V} &=&  \frac{\Lambda_{W}  }{x^{2} }
\Big[2 + 3 x^{2} + 2\big(1 - 4\alpha_{\pm}\big) x^{2\beta_{\pm}}\Big]\nb\\
& & + \frac{\Lambda_{W} }{x^{8(1-\alpha_{\pm})} }
\Big\{\big[8\xi \alpha_{\pm}^{2} - 8(1-\xi) \big(1 - \alpha_{\pm}\big) + 3\big] \nb\\
& &
+ 2\big[4(1-\xi)(1- \alpha_{\pm}) - 1\big] x^{2(1-2\alpha_{\pm})} \nb\\
& & 
+ (1-2\xi)  x^{4(1-2\alpha_{\pm})}\Big\},\nb\\
\eqn
where $\Delta$ is given by Eq. (\ref{3.17}), and $\Delta' \equiv d\Delta/dx$. The quantity
$v$, defined by
 \bq
 \lb{a.5} 
 J^{i} \equiv - N\frac{\delta{\cal{L}}_{M}}{\delta N_{i}} = e^{-(\mu+\nu)}\big(v, 0, 0\big),
 \eq
which is related to   heat flow \cite{GPW}, is given by
\bqn
\lb{a.6} 
v &=&  \frac{\Lambda_{W} e^{2(\mu-\nu)}}{32\pi G x^{2}\Delta^{2}}
\Bigg\{\xi x\Big[2x\Delta\Delta'' +   x {\Delta'}^{2} - 2 \beta_{\pm}  \Delta\Delta'\Big] \nb\\
& & + 4\Delta\Big[\xi x\Delta' -   2(1-\xi)\beta_{\pm}\Delta\Big] - 8 \xi\Delta^{2}\Bigg\}.
\eqn

On the other hand, the corresponding stress part $\tau^{ij}$ defined by,
\bq
\lb{a.7}
\tau^{ij} = {2\over \sqrt{g}}{\delta \left(\sqrt{g}
 {\cal{L}}_{M}\right)\over \delta{g}_{ij}},
 \eq
can be written in the form, 
\bq
\lb{a.8}
\tau_{ij}  =  e^{2\nu}p_{r} \delta^{i}_{r} \delta^{j}_{r}
 + r^{2}p_{\theta}\Omega_{ij},
  \eq
where $\Omega_{ij} \equiv    \delta_{i}^{\theta} \delta_{j}^{\theta}  +
\sin^{2}\theta \delta_{i}^{\phi} \delta_{j}^{\phi}$, and 
\bqn
\lb{a.9}
p_{r} &=&  \frac{\Lambda_{W} e^{2(\mu-\nu)}}{64\pi G x^{2}\Delta^{2}}
\Bigg\{\xi x\Big[4x\Delta\Delta'' -3  x {\Delta'}^{2} +4 \beta_{\pm}  \Delta\Delta'\Big] \nb\\
& & + 8\Delta\Big[x\Delta' -   2(1-\xi)\beta_{\pm}\Delta\Big] + 8(1-4 \xi)\Delta^{2}\Bigg\}\nb\\
& & -  \frac{ e^{-2\nu}}{8\pi G} F_{rr},\nb\\
p_{\theta} &=&  \frac{\Lambda_{W} e^{2(\mu-\nu)}}{16\pi G x^{2}\Delta^{2}}
\Bigg\{(1-\xi) x^{2}\Delta\Delta'' + \frac{1}{4}\xi x^{2} {\Delta'}^{2}  \nb\\
& & + (1-\xi) \beta_{\pm}  x\Delta\Delta' + 2 (1-2\xi)\big(x \Delta'+ \beta_{\pm}\Delta\big)\Delta\Bigg\}\nb\\
& & +  \frac{\Lambda_{W}}{8\pi G x^{2}} F_{\theta\theta},
\eqn
 with 
\bqn
\lb{a.10}
F_{rr} &=&  \frac{\Lambda_{W}  }{2 x^{2}}\Big[2 - 2 x^{-2\beta_{\pm}} -  3 x^{2(1-\beta_{\pm})} \Big] \nb\\
& &  - \frac{\Lambda_{W}  }{4 x^{4-2\beta_{\pm}}}\Bigg[ \big(19 - 32\xi\big) \nb\\
& & + 4(3-5\xi)\big(2 - 3\beta_{\pm}\big) \beta_{\pm}\nb\\
& & - 2\big(7 -12\xi\big)x^{-2\beta_{\pm}} - \big(5 - 8\xi\big)x^{-4\beta_{\pm}}\Bigg],\nb\\
F_{\theta\theta} &=&   \frac{3}{2} x^{2} - \beta_{\pm} x^{2\beta_{\pm}} + \frac{1}{2x^{2 - 4\beta_{\pm}}}\Bigg\{(1+2\xi)\nb\\
& &  + 2(11 - 16\xi)\beta_{\pm} - 2\xi\big(7 - 6\beta_{\pm}\big)\beta_{\pm}^{2} \nb\\
& & - 2\big(1 - \beta_{\pm}\big)x^{-2\beta_{\pm}} + (1-2\xi) x^{-4\beta_{\pm}}\Bigg\}. ~~
\eqn
   
Clearly, to interpret the above as an anisotropic fluid with heat flow, some energy conditions \cite{HE72}
might need to be imposed. Since in this paper we are mainly concerned with the nature of singularities,
we shall not study these conditions here.
 

\end{document}